\documentclass[aps,prx,reprint,superscriptaddress]{revtex4-2}
\usepackage{amsmath}
\usepackage{graphicx}
\usepackage{psfrag}
\usepackage{amssymb}
\usepackage{color}
\usepackage{hyperref}
\usepackage{ulem}

\begin{document}


\title{Two-component electronic phase separation in the doped Mott insulator Y$_{1-x}$Ca$_{x}$TiO$_{3}$}    
\author{S. Hameed}
\thanks{Corresponding authors: hamee007@umn.edu, greven@umn.edu}
\affiliation{School of Physics and Astronomy, University of Minnesota, Minneapolis, MN 55455, U.S.A.}%
\author{J. Joe}
\affiliation{School of Physics and Astronomy, University of Minnesota, Minneapolis, MN 55455, U.S.A.}%
\author{D. M. Gautreau}
\affiliation{School of Physics and Astronomy, University of Minnesota, Minneapolis, MN 55455, U.S.A.}%
\affiliation{Department of Chemical Engineering and Materials Science, University of Minnesota, Minneapolis, Minnesota 55455, USA.}%
\author{J. W. Freeland}
\affiliation{X-ray Science Division, Argonne National Laboratory, Argonne, IL 60439, USA.}
\author{T. Birol}
\affiliation{Department of Chemical Engineering and Materials Science, University of Minnesota, Minneapolis, Minnesota 55455, USA.}%
\author{M. Greven}
\thanks{Corresponding authors: hamee007@umn.edu, greven@umn.edu}
\affiliation{School of Physics and Astronomy, University of Minnesota, Minneapolis, MN 55455, U.S.A.}%

\widetext
\date{\today}

\begin{abstract}
One of the major puzzles in condensed matter physics has been the observation of a Mott-insulating state away from half-filling. The filling-controlled Mott insulator-metal transition, induced via charge-carrier doping, has been extensively researched, but its governing mechanisms have yet to be fully understood. Several theoretical proposals aimed to elucidate the nature of the transition have been put forth, a notable one being phase separation and an associated percolation-induced transition. In the present work, we study the prototypical doped Mott-insulating rare-earth titanate YTiO$_3$, in which the insulating state survives up to a large hole concentration of 35\%. Single crystals of Y$_{1-x}$Ca$_x$TiO$_3$ with $0 \leq x \leq 0.5$, spanning the insulator-metal transition, are grown and investigated. Using x-ray absorption spectroscopy, a powerful technique capable of probing element-specific electronic states, we find that the primary effect of hole doping is to induce electronic phase separation into hole-rich and hole-poor regions. The data reveal the formation of electronic states within the Mott-Hubbard gap, near the Fermi level, which increase in spectral weight with increasing doping. From a comparison with DFT+$U$ calculations, we infer that the hole-poor and hole-rich components have charge densities that correspond to the Mott-insulating $x = 0$ and metallic $x \sim 0.5$ states, respectively, and that the new electronic states arise from the metallic component. Our results indicate that the hole-doping-induced insulator-metal transition in Y$_{1-x}$Ca$_x$TiO$_3$ is indeed percolative in nature, and thus of inherent first-order character.

\end{abstract}
\maketitle



\section{Introduction}
Mott insulator-metal transitions (IMT) are ubiquitous in condensed matter physics and have been the basis of many decades of experimental and theoretical research \cite{Imada1998}. ln the theoretically ideal case of a lattice model with a single electronic orbital per site, the Mott insulating state exists only at half-filling (one electron per site), as on-site electron-electron Coulomb repulsive interactions localize conduction electrons in a half-filled band. However, experiments on several transition-metal oxides have shown that the Mott-insulating state can survive up to nonzero charge-carrier doping \cite{Garcia-Munoz1995,Katsufuji1997,Tokura2006,Fujioka2005}. The governing mechanisms of this filling-controlled IMT is an important longstanding question \cite{Imada1998}.

The rare-earth titanates RTiO$_{3}$ (R is a rare earth atom) are prototypical Mott insulators in which such a filling-controlled IMT can be induced $via$ hole doping, which effectively changes the 3$d$ band-filling \cite{Mochizuki2004}. These perovskites exhibit a relatively simple pseudocubic structure and can be considered model systems with which to explore the physics of the IMT. In contrast to the well-known Ti$^{4+}$ systems SrTiO$_{3}$ and BaTiO$_3$, the Ti$^{3+}$ ion in RTiO$_{3}$ has a spin-$\frac{1}{2}$ $3d^1$ electronic configuration. Recent years have seen renewed interest in the RTiO$_{3}$ materials, particularly concerning the electronic and magnetic structure of the insulating state \cite{Yue2020,Hameed2021} and the nature of the Mott transition \cite{Yee2015,Kurdestany2017,Kim2017,German2018}. The rare-earth ion R in RTiO$_{3}$ is an important factor in determining the ground state magnetic and electronic properties, primarily because it controls the Ti-O-Ti bond angle (and hence the electron bandwidth) which is less than 180$^o$ due to the distorted perovskite structure \cite{Mochizuki2004}. 

Depending on the rare-earth ion, the doping level at which the system becomes metallic varies widely, from $x_c = 0.05$ in La$_{1-x}$Sr$_{x}$TiO$_{3}$ \cite{Tokura1993La} to $x_c = 0.15$ and $x_c = 0.35$, respectively, in Nd$_{1-x}$Ca$_{x}$TiO$_{3}$ \cite{Ju1994} and Y$_{1-x}$Ca$_{x}$TiO$_{3}$ \cite{Tokura1993}. Prior experimental studies attributed these differences to the R-dependent electronic bandwidth \cite{Katsufuji1997}. Recent theoretical work based on thermodynamic arguments, with specific focus on RTiO$_3$, suggests that the doped system electronically phase-separates into metallic hole-rich regions and Mott-insulating hole-poor regions, and therefore points to a percolative nature of the transition and an underlying bandwidth-control mechanism \cite{Yee2015,Kurdestany2017}. The hole concentration in the hole-rich puddles was found to strongly depend on the electron-lattice coupling strength and the electronic bandwidth \cite{Yee2015,Kurdestany2017}. Such electronic phase separation has long been predicted even in simple Hubbard models \cite{Gehlhoff1996, Werner2007}. Signatures of electronic phase separation have indeed been observed experimentally in Seebeck and magnetization measurements of bulk LaTiO$_{3+\delta}$ \cite{Zhou2005b}. In bulk Y$_{1-x}$Ca$_{x}$TiO$_{3}$, low-temperature structural phase separation has been reported in the doping range $x = 0.37-0.41$ that is in the vicinity of the IMT \cite{Kato2002a,Tsurui2004,German2018}, which is likely a consequence of possible electronic phase separation. However, no coupling of the IMT to a specific lattice symmetry has been observed in films of Sm$_{1-x}$Sr$_{x}$TiO$_{3}$ \cite{Kim2017}.

We aim to elucidate the nature of this filling-controlled Mott transition by focusing on Y$_{1-x}$Ca$_{x}$TiO$_{3}$, since the insulating state is stabilized over an extended doping range in this system. An interesting property of YTiO$_3$ (YTO) is its ferromagnetic-insulating ground state as opposed to an antiferromagnetic-insulating ground state that one would naively expect based on the Hubbard model. This property has been attributed to the orbital-order that is known to exist in YTO \cite{Mochizuki2004}, which may also be relevant for the physics of the IMT. In addition to the hole-doping-induced IMT, Y$_{1-x}$Ca$_{x}$TiO$_{3}$ exhibits a temperature-induced IMT at intermediate doping levels \cite{Tokura1993,Kato2002a,Tsurui2004}.
Through x-ray absorption spectroscopic (XAS) measurements of single crystals at the Ti $L$- and O $K$-edges, we find that the primary effect of hole doping is electronic phase separation into hole-rich metallic and hole-poor Mott-insulating regions. Our data furthermore reveal the formation of in-gap states near the Fermi level. These states exhibit an increasing spectral weight with increasing doping and decreasing temperature, presumably due to an increasing volume fraction of the metallic regions. Based on a comparison with DFT+U calculations, we argue that the hole-poor component has zero hole-density $p=0$, and thus corresponds to the undoped Mott-insulating state of YTO ($x = 0$), whereas the hole-rich component has a hole-density of $p\sim0.5$ that corresponds to the metallic state of Y$_{0.5}$Ca$_{0.5}$TiO$_3$. Our results strongly indicate a percolative, first-order nature of the doping-induced IMT, consistent with theory \cite{Yee2015,Kurdestany2017}.


\section{Methods}
Y$_{1-x}$Ca$_{x}$TiO$_{3}$ single crystals were melt-grown with the optical floating-zone technique \cite{Komarek2009}. A single crystal with $x = 1$ was purchased from PI-KEM Ltd.
Laue x-ray diffraction was used to confirm single crystallinity.

Resistivity measurements were carried out with a Quantum Design, Inc. PPMS Dynacool system. A Keithley 2612B sourcemeter was used in the four-wire configuration to source the current and measure the voltage. The samples were cut into a square shape with an $ab$-plane surface, the $a$- and $b$-axes being the sides of the square. The sample surfaces were polished to a surface roughness of $\sim$ 0.3 $\mu$m using polycrystalline diamond suspension, and contacts were made by aluminum-wire bonding onto sputtered gold pads or by directly bonding aluminum wire onto the sample surface. All samples were measured in a van der Pauw geometry (vdP) except $x = 0.35$ where a stripe geometry was used.

XAS measurements were carried out at beam-line 4-ID-C of the Advanced Photon Source, Argonne National Lab. The monochromator resolution was set at 0.1 eV. Using a SrTiO$_3$ energy reference recorded simultaneously, the spectra energy scales were aligned to within 0.1 eV. The samples for XAS measurements were picked from the same section of the growth as the transport samples. The sample surfaces were cut along the $ab$-plane and polished with the same protocol as the transport samples. 
The x-rays were incident at an angle of 15$^\text{o}$ with respect to the $c$-axis. Although XAS measurements were performed in both Total Fluorescence Yield (TFY) and Total Electron Yield (TEY) modes, the TEY mode data featured a spurious Ti$^{4+}$ contribution, as the surface of RTiO$_{3}$ easily oxidizes to form Ti$^{4+}$ (see \cite{SM} for a comparison to CaTiO$_{3}$ (CTO; $x = 1$) spectra illustrating this). Therefore, all data reported here were obtained in the bulk-sensitive TFY mode.

DFT calculations were performed using the projector augmented wave approach as implemented in the Vienna Ab Initio Simulation Package (VASP) \cite{Kresse1993,Kresse1996CMS,Kresse1996PRB}. Calculations on YTiO$_3$ were performed using the PBEsol exchange-correlation functional \cite{Perdew2008}, a plane wave cutoff of 550 eV, and a $4\times4\times4$ Monkhorst-Pack grid for $2\times2\times2$ supercells. In order to properly reproduce the local magnetic moments on the Ti ions, these calculations used the rotationally-invariant LSDA+$U$ scheme \cite{Dudarev1998} with $U=4~$eV. We initialized the system in a collinear ferromagnetic state, in which the system remained after self-consistency was achieved. We calculated the projected density of states for the cases where zero to four holes have been introduced into the system. Since the $2\times2\times2$ supercell contains eight Ti atoms, these hole concentrations correspond to $p = 0$ to 0.5 holes per Ti ion. The crystal structure was assumed to be that of the orthorhombic (\textit{Pnma}) YTiO$_3$ throughout.

\section{Results}

\begin{figure}
\includegraphics[width=0.4\textwidth]{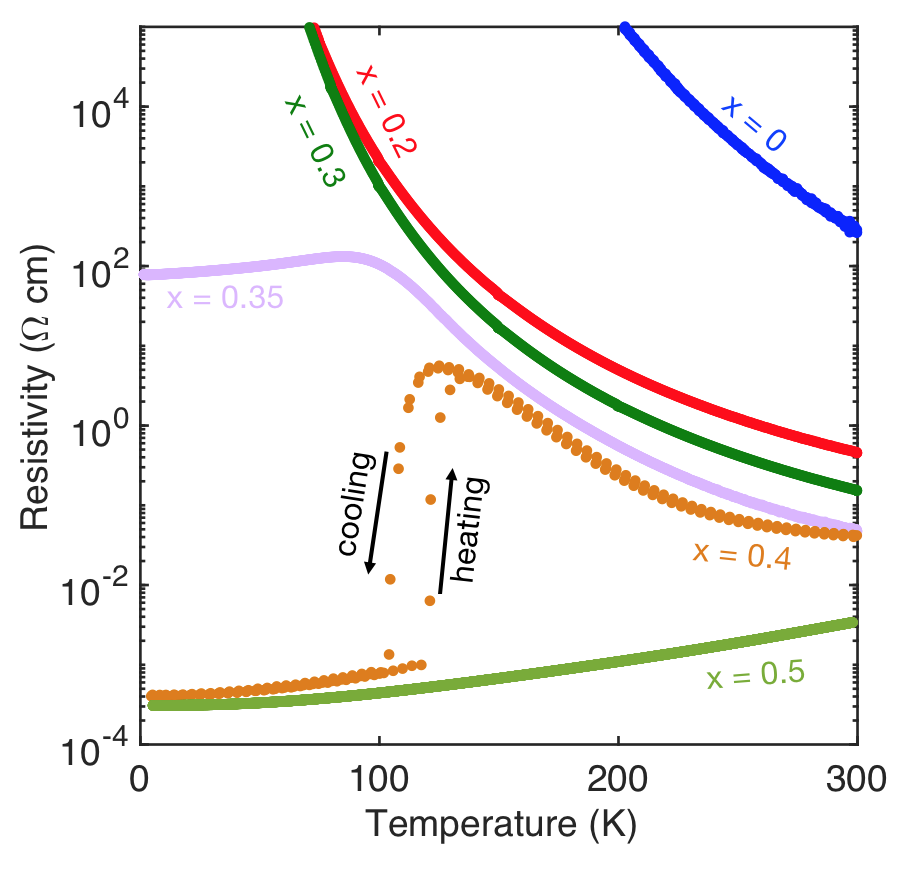}
\caption{Doping and temperature dependence of resistivity in Y$_{1-x}$Ca$_{x}$TiO$_{3}$.}
\label{fig:Resistivity}
\end{figure}

\begin{figure}
\includegraphics[width=0.5\textwidth]{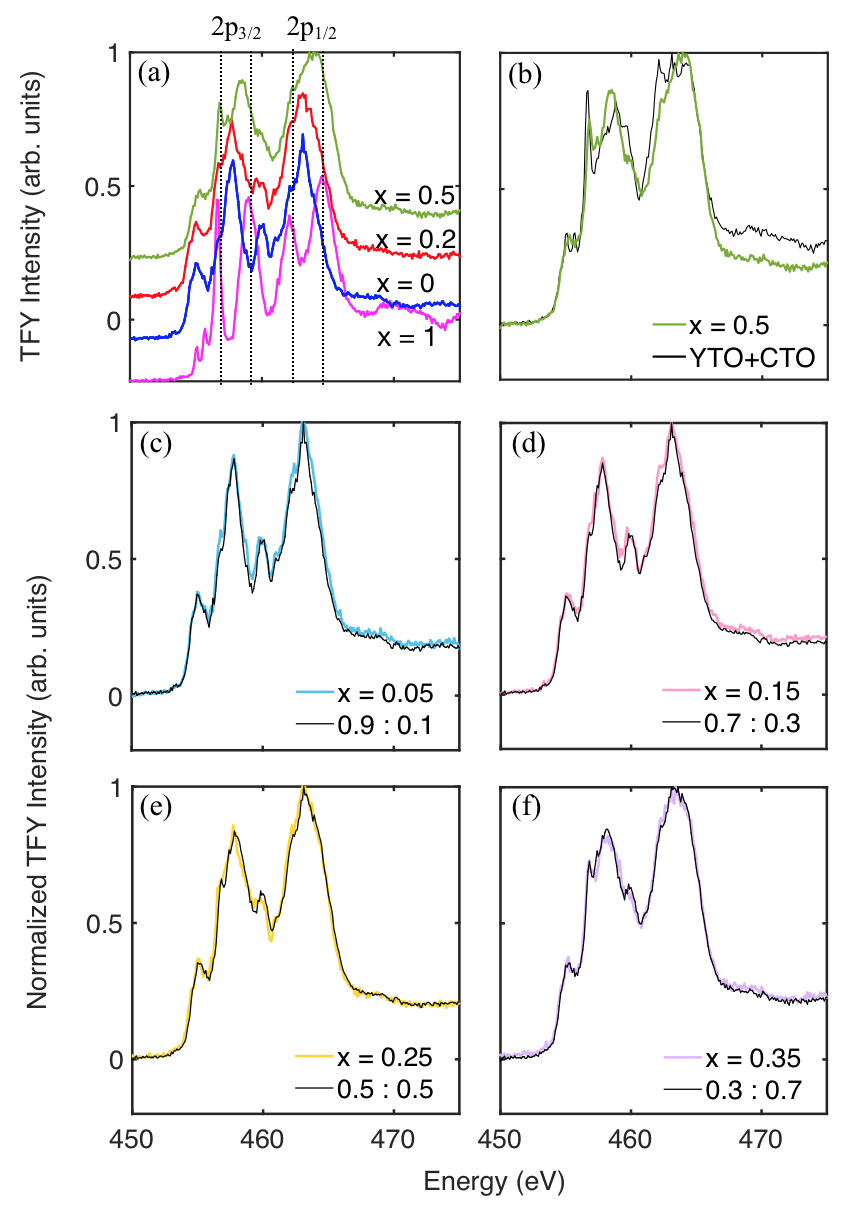}
\caption{(a) Ti $L$-edge XAS spectra for Y$_{1-x}$Ca$_{x}$TiO$_{3}$, including both YTO ($x=0$) and CTO ($x=1$).
(b) XAS spectra for $x= 0.5$ compared with an equally-weighted linear combination of the YTO and CTO spectra in (a). (c)-(f) XAS spectra at four doping levels, compared with a linear combination of the $x=0$ and $x=0.5$ spectra (ratio of weights as indicated). All spectra were obtained at 15 K and normalized to the highest peak intensity.}
\label{fig:XrayTiL}
\end{figure}

Figure~\ref{fig:Resistivity} shows the temperature and doping dependence of the resistivity. The samples in the $x = 0-0.3$ range are insulating, whereas the $x = 0.5$ system is metallic below room temperature. A temperature-induced IMT can be observed for $x = 0.35$ and $0.4$, with thermal hysteresis for $x = 0.4$. Prior structural studies of samples with $x \leq 0.38$ showed that, whereas the system adopts an orthorhombic crystal structure at 300 K, it transitions to a monoclinic phase at $\sim$ 200 K \cite{Tsurui2004}. Upon decreasing the sample's temperature further, a structural separation into a monoclinic phase and an orthorhombic phase was observed for $x = 0.37 - 0.41$ and the low temperature orthorhombic phase was argued to be metallic in character \cite{Kato2002a,Tsurui2004,German2018}. Such inhomogeneity manifested itself in our transport measurements as well. For $x = 0 - 0.3$, the resistivity exponentially increases with decreasing temperature and crosses the measurement limit at low temperatures. No significant anisotropy was observed in this doping range for resistance measured with current along the $a$- and $b$-axes in the vdP geometry. For $x = 0.35$, measurement of the low-temperature resistance was not possible in the vdP geometry. However, a stripe geometry enabled measurements down to low temperatures. This is likely because in the vdP geometry, the resistance measurement relies on the point current contacts being able to distribute the current through the sample volume to reach the voltage measurement contacts. In this case, measurement in the vdP geometry is hindered by a large volume fraction of the insulating phase in $x = 0.35$ at low temperatures. A stripe geometry circumvents this issue by providing a larger area for the current and voltage contacts. For $x = 0.4$ and $0.5$, the resistances could be measured in the vdP geometry to low temperatures, which may be due to the larger metallic volume fractions present at these compositions. We note that significant resistivity anisotropy has been reported previously near $x = 0.38$, likely due to structural phase separation \cite{Tsurui2004}. Hence, for $x = 0.35 - 0.5$ in Fig.~\ref{fig:Resistivity}, we measured the resistivity with current along the $b$-axis, whereas for $x = 0 - 0.25$ we used the usual vdP resistivity for the $ab$-plane. Further analysis of the resistivity data is presented in \cite{SM}. The results are in good agreement with prior reports \cite{Tokura1993,Kato2002a,Tsurui2004}.

\begin{figure*}
\includegraphics[width=\textwidth]{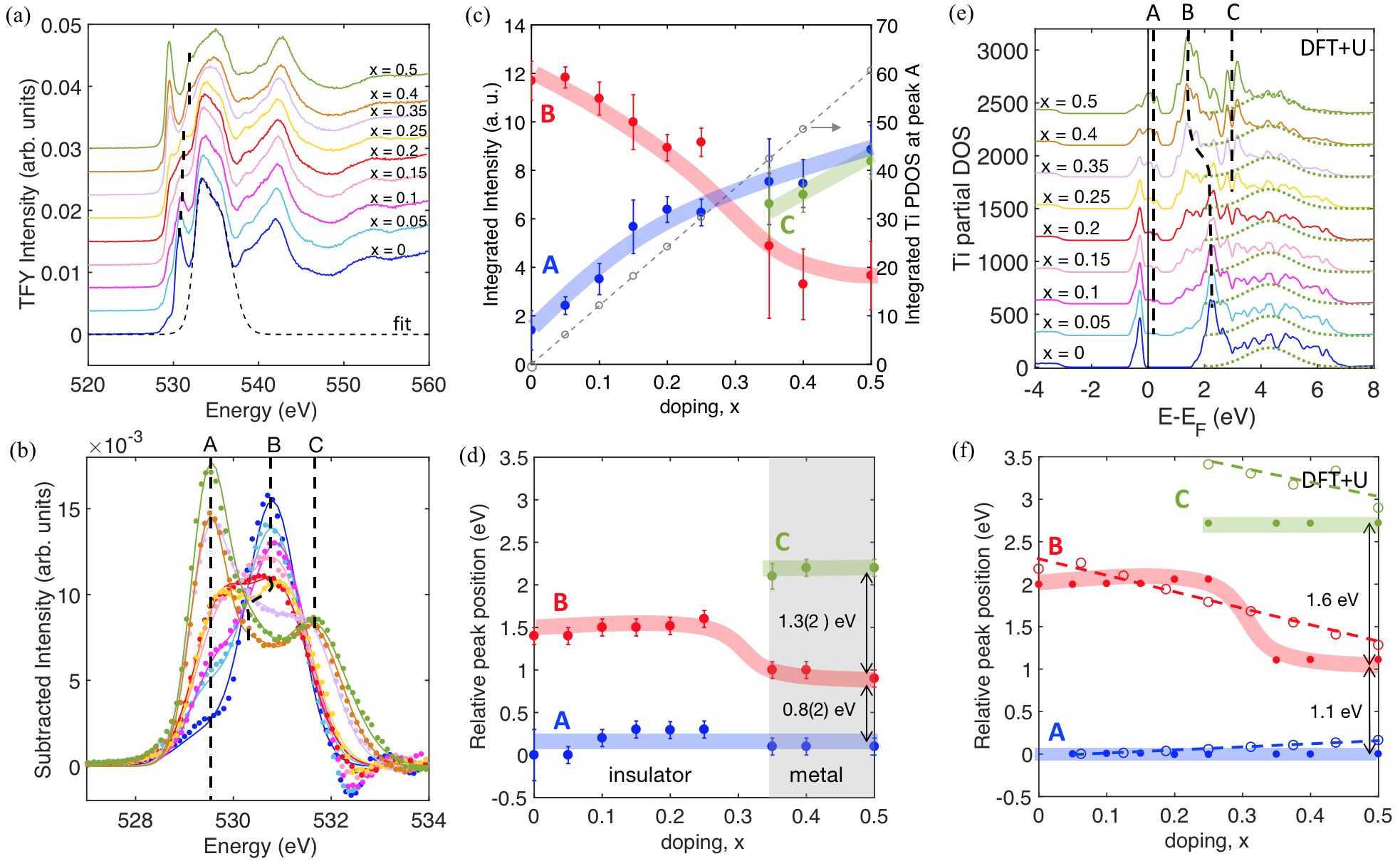}
\caption{(a) Doping dependence of O $K$-edge XAS at 15 K. The spectra are shifted vertically for clarity. In order to extract the pre-edge peaks, we fit the data around 535 eV to two gaussians (fits shown for $x=0$ as a dashed line) and subsequently subtracted the fit result from the data to obtain the result in (b). 
Black markers are guides to the eye and track the higher-energy component of the pre-edge. 
(b) Pre-edge peaks at the O $K$-edge. 
The lines are the results of fits to two or three gaussians, as described in the text. The vertical black dashed lines track the distinct peak positions.
(c,d) Energy-integrated intensities and peak positions, respectively, of the gaussian peaks in (b). The thick lines are guides to the eye.
(e) Doping dependence of the Ti PDOS obtained from DFT+$U$ calculations. The PDOS at $x = 0$ ($p = 0$) and $x = 0.5$ ($p = 0.5$) are obtained directly from the calculations, whereas those at intermediate doping levels are obtained by taking a weighted average of $x = 0$ and $x = 0.5$ PDOS, with the weights chosen in accordance with Fig.~\ref{fig:XrayTiL}. The results of the calculation are convolved with a gaussian function of width 0.1 eV (full-width-at-half-maximum) to mimic the experimental resolution. 
The vertical black dashed lines track the three distinct peaks that correspond to the pre-edge peaks observed in (b). The dashed green lines are gaussian fits to the broad high-energy component at $x = 0.5$, used to highlight the transfer of PDOS from the high-energy region to low-energy region, forming peak C. 
(f) Doping dependence of the peak positions obtained from (e) (filled symbols); the position for A is chosen as the midpoint of the nonzero-PDOS region just above the Fermi level, as XAS only sees states above the Fermi level. The thick lines are guides to the eye. The integrated Ti PDOS for the region corresponding to A (above the Fermi level) is presented on the right axis in (c), with a dashed line as a guide to the eye. The peak positions in (d) and (f) are plotted relative to the peak A position at $x = 0.05$. The open symbols in (f) are the peak positions obtained from Ti PDOS, directly calculated by DFT+$U$, without considering phase separation. In displaying this, we assume $x = p$. The dashed lines are guides to the eye. A clear qualitative difference (continuous vs. discontinuous) in the peak B position is observed between the data and calculation, in this case.}
\label{fig:XrayOK}
\end{figure*}


In order to clarify the nature of this IMT, we carried out detailed XAS measurements.
Ti $L$-edge XAS is a direct probe of the valence state of the Ti ion, which is expected to undergo changes with hole doping.
Figure~\ref{fig:XrayTiL}(a) shows the doping dependence of the Ti $L$-edge at a few selected doping levels. A reference CTO spectrum is added for comparison. The spectra for YTO and CTO exhibit four features that correspond to the transitions from Ti 2$p_{1/2}$ and 2$p_{3/2}$ into the t$_{2g}$ and e$_{g}$ orbitals. Upon substitution of Y with Ca in YTO, the main peaks are seen to shift to higher energies, and new features appear at the peak energies corresponding to CTO, due to a partial conversion of Ti$^{3+}$ to Ti$^{4+}$. However, a comparison of the $x=0.5$ spectrum with an equally-weighted linear combination of YTO ($x=0$) and CTO ($x=1$) spectra in Fig.~\ref{fig:XrayTiL}(b) shows little agreement, which indicates that the effect of Ca-doping is not merely a volume-wise separation into Ti$^{3+}$ and Ti$^{4+}$ regions (c.f. \cite{Mizumaki2004,Ulrich2008}). In Figs.~\ref{fig:XrayTiL}(c)-(f), we instead compare the measured Ti $L$-edge spectra at different doping levels to weighted averages of $x=0$ and $x=0.5$ spectra. We assume a simple, linear volume fraction change of the two different kinds of Ti valence-states: e.g.,  70\% that of $x=0$ and 30\% that of $x=0.5$ for $x=0.15$.
The excellent agreement between the calculated and measured spectra indicates that the effect of hole-doping can be viewed as electronic phase separation into regions with two different Ti valence-states (and hence hole-concentrations) resembling that of the Mott-insulating $x = 0$ and metallic $x = 0.5$ states, with a simple linear-in-$x$ volume-fraction dependence. 

This is in excellent agreement with the proposed theoretical picture \cite{Yee2015,Kurdestany2017}, where any nonzero hole-concentration $p>0$ is argued to lead to electronic phase separation into regions with $p = 0$ and a nonzero hole concentration $p = p^*$, where $p^*$ is the hole-concentration required to form a single-phase metal. Percolation of the $p = p^*$ regions at some critical average hole-concentration $p_c$ then leads to the IMT. Let us designate the Ca-concentration required to attain a hole-concentration $p^*$ (and hence a single-phase metal) as $x^*$. From the XAS data alone, we can not conclude that $x^* \sim 0.5$, as the Ti $L$-edge spectra at any intermediate doping level may be represented as a linear combination of $x = 0$ and some sufficiently large $x$. However, the transport data (Fig.~\ref{fig:Resistivity}) provide the constraint that $x^*$ can not be much less than $0.5$, as the metallic phase emerges only at $x_c = 0.35$ and a temperature-dependent (hysteretic) IMT is observed at $x=0.4$. The latter is distinct from the simple metallic behavior at $x=0.5$ and likely a reflection of two-phase coexistence. 


With the goal to further understand the effects of the electronic phase separation on the electronic states, we determined the doping and temperature dependence of the O $K$-edge spectra. O $K$-edge XAS is a direct probe of the unoccupied density of states in the system, and hence enables a comparison to DFT+$U$ calculations. Figure~\ref{fig:XrayOK}(a) shows the doping dependence of the O $K$-edge spectra. Three main features can be observed for YTO, corresponding to transitions from O 1$s$ to unoccupied O 2$p$ states hybridized with Ti $3d$ ($\sim$ 530 eV), Y 4$d$/Ca 3$d$ ($\sim$ 535 eV) and Ti $4sp$ ($\sim$ 542 eV), respectively \cite{Suntivich2014}. While small shifts can be discerned in the O 2$p$-Y $4d$/Ca $3d$ and O 2$p$-Ti $4sp$ regions, major changes are observed in the O 2$p$-Ti 3$d$ pre-edge region. In order to visualize this better, we removed the XAS spectra above the pre-edge region. This was achieved by fitting the data around 535 eV to a pair of gaussians, as shown for YTO in Fig.~\ref{fig:XrayOK}(a), where the fit was in the energy range 532.5 - 536.5 eV. The peak positions of the gaussians were subsequently fixed at nonzero doping, as the higher-energy region of the total DOS does not show significant shifts in energy with hole doping (see \cite{SM}). Figure.~\ref{fig:XrayOK}(b) shows the pre-edge peak signal. For $x\leq0.25$, two distinct peaks A (lower energy) and B (higher energy) are resolved. At higher doping ($x = 0.35$, 0.4 and 0.5), an additional peak appears at an energy higher than peak B and combines with the Y $4d$/Ca $3d$ region. Therefore, for $x \geq 0.35$, a smaller energy range 533.5 - 536.5 eV was used for the fit. 
In order to obtain a quantitative description of the pre-edge peaks, we fit the data to two and three gaussian peaks for $x\leq0.25$ and $x\geq0.35$, respectively, and extracted the peak positions and energy-integrated intensities as shown in Fig.~\ref{fig:XrayOK}(c,d). We discern a ``jump'' in peak B position between $x = 0.25$ and 0.35, and interpret this as a cross-over from $p = 0$ to $p = p^*$ dominated spectra, as explained below.

We compare the data with the calculated Ti partial density of states (PDOS), as major changes in the low-energy region are observed only in the Ti PDOS (see \cite{SM}). 
For illustrative purposes, we assume $x^* = 0.5$ in the comparison with the DFT+$U$ calculations. Since it is known that $p = x$ in the doping range $x = 0.5 - 1$ \cite{Tokura1993}, we assume in calculating the DOS that $x = 0.5$ has a hole concentration of $p = 0.5$. At each doping level, a weighted average of the DFT-calculated Ti PDOS for $p = 0$ (corresponding to $x = 0$) and $p = 0.5$ (corresponding to $x = 0.5$) is used, with weights in accordance with Fig.~\ref{fig:XrayTiL}. The results thus obtained are plotted in Fig.~\ref{fig:XrayOK}(e). 
From a comparison with Fig.~\ref{fig:XrayOK}(c), it can be clearly seen that the increasing intensity of peak A with doping in the O $K$-edge data is due to an increasing Ti PDOS at and right above the Fermi level. The appearance of three distinct peaks at high doping in the O $K$-edge data can be clearly traced to three distinct peaks in the Ti PDOS, with peak C being discernible only at high doping, where it emerges out of the broad high-energy component. Moreover, the doping dependence of the peak positions and their relative energies in the Ti PDOS in Fig.~\ref{fig:XrayOK}(f) show great agreement with the experimental results in Fig. \ref{fig:XrayOK}(d). A crossover is seen around $x = 0.25$ in the Ti PDOS, where peak B has approximately equal contributions from the $p = 0$ and $p = 0.5$ DOS, and hence no distinct peak is visible. The ``jump'' in peak B position between $x = 0.25$ and $x = 0.35$ in Fig.~\ref{fig:XrayOK}(d) is thus associated with a crossover from $p = 0$ to some $p = p^*$ dominated spectra. 
Given the absence of hysteresis and of a thermal phase transition in the $x = 0.5$ resistivity data (Fig.~\ref{fig:Resistivity}) as well as the excellent agreement between the XAS data for $x = 0.5$ and theoretical results for $p=0.5$, the hole-rich metallic phase likely has a hole-concentration $p^*\sim x^*\sim 0.5$. Of course, exact agreement is not expected, as the oxygen core-hole potential and finite lifetime of the O $1s$ are ignored in our DFT+$U$ calculations. For comparison, we also present the peak positions extracted directly from DFT+$U$ calculations, without considering any phase separation, in Fig.~\ref{fig:XrayOK}(f) (See \cite{SM} for the detailed DOS). Whereas the experimentally-obtained peak B position shows a discontinuity between $x = 0.25$ and 0.35, the theoretical peak position exhibits a qualitatively different, continuous change. This further supports the electronic phase-separation scenario.

The spacing between the A and B peaks in Fig.~\ref{fig:XrayOK}(d) is 1.3(2) eV, 
comparable to the Mott-Hubbard gap of about 1 to 1.5 eV that has previously been observed in YTiO$_{3}$ \cite{Taguchi1993,Yue2020}. Given that peak B arises from the upper Hubbard band \cite{Arita2007a}, the emergence and enhancement of the lower-energy peak A at the O $K$-edge indicates the formation of in-gap states of mixed Ti $3d$ and O $2p$ character similar to that observed in the high $T_c$ cuprates upon hole doping \cite{Chen1991}. Note that the nonzero intensity of peak A at $x = 0$ in the data could be due to small unintentional doping in the form of oxygen off-stoichiometry. Peak A (which is the peak that is closest to the Fermi level) shows no discontinuity in intensity and position at $x_c = 0.35$, indicative of a smooth cross-over from an insulator to a metal associated with a gradual increase in volume fraction of the hole-rich metallic phase. This is in agreement with previous optical-conductivity data, which also revealed a gradual shift of the spectral intensity from the Mott-Hubbard gap excitations to the inner-gap region with doping \cite{Taguchi1993}. 
The spacing between peak A and peak C is 2.1(2) eV, in agreement with a prior XAS study of an $x = 0.39$ single crystal \cite{Arita2007a}. Note that unnormalized data are used to extract the pre-edge region in Fig.~\ref{fig:XrayOK}(b). We show in \cite{SM} that normalization by the intensity at 560 eV leaves the qualitative results unchanged. 

While new states at the Fermi level are clearly observed at the O $K$-edge already at low doping, bulk metallicity arises only at $x = 0.35$. This further supports the interpretation of the Ti $L$-edge data that a small volume fraction of the metallic hole-rich $p^* \sim 0.5$ phase is present already at low doping.

\begin{figure}
\includegraphics[width=0.4\textwidth]{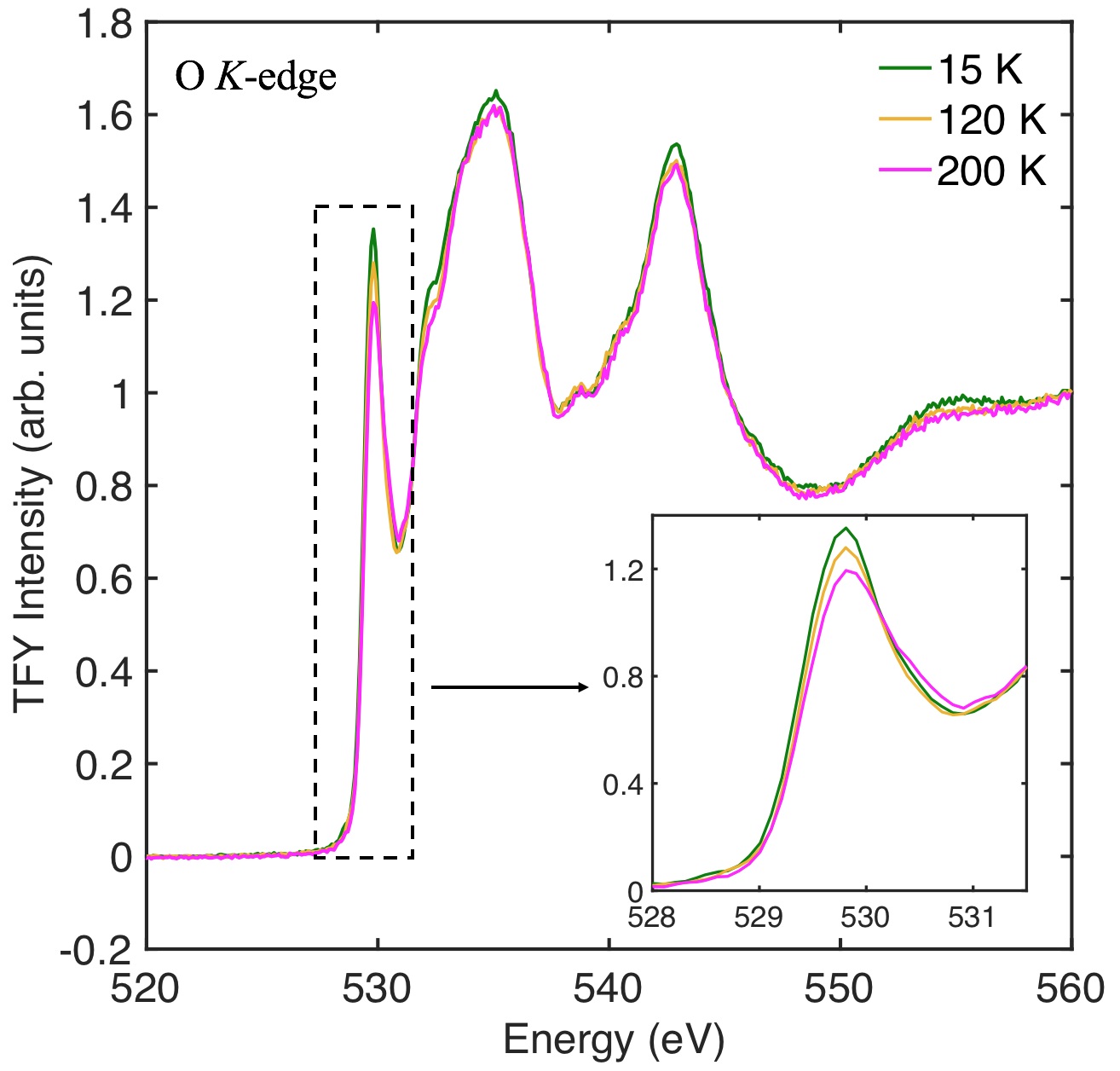}
\caption{O $K$-edge XAS spectra at $x = 0.4$ at different temperatures spanning the IMT. The data were obtained on heating. The inset shows a magnified view of the pre-edge region. The spectra have been normalized to the intensity at 560 eV.}
\label{fig:XrayOKTdep}
\end{figure}

Given the temperature-induced IMT observed at $x = 0.35$ and $x = 0.4$ (Fig.~\ref{fig:Resistivity}), we also measured the temperature dependence of the O $K$-edge spectra for $x = 0.35$ to 0.5. Figure~\ref{fig:XrayOKTdep} shows the O $K$-edge spectra for $x = 0.4$ at three different temperatures: 15 K, well inside the metallic phase; 120 K, near the peak in resistivity (IMT temperature) in Fig.~\ref{fig:Resistivity}; and 200 K, well above the IMT. The spectra are normalized to the intensity at 560 eV. A magnified view of the pre-edge peak region is shown in the inset. It can be clearly seen that the primary effect of decreasing temperature is an increase in the pre-edge peak intensity (peak A in Fig.~\ref{fig:XrayOK}(b)) which is similar to the increase in intensity of peak A that we observe with increased doping in Fig.~\ref{fig:XrayOK}(c). No shift of the pre-edge peak position with temperature is observed within the experimental resolution. Such an enhancement in the pre-edge peak intensity with decreasing temperature is also observed at $x = 0.35$ and 0.5 (see \cite{SM}). This is consistent with a previous XAS study of a sample with $x = 0.39$ that was attributed to increased O 2$p$-Ti 3$d$ hybridization with decreasing temperature \cite{Arita2007a}. From our interpretation of the O $K$-edge data, we believe that this intensity change could be due to a strong temperature dependence of the Ti-O-Ti bond angle (and hence the electronic bandwidth) and/or electron-lattice coupling, both of which can significantly affect the value of $p^*$ \cite{Yee2015,Kurdestany2017}. This would then lead to a change in volume fraction of the metallic $p = p^*$ phase with temperature, and hence of the pre-edge peak intensity. Alternatively, the intensity enhancement could also be due to the temperature-dependent changes in lattice symmetry that have been reported in this doping range \cite{Kato2002a,Tsurui2004,German2018}. Note that no temperature dependence is observed within experimental error in the Ti $L$-edge (see \cite{SM}), indicating the absence of a change in the average Ti valence-state with temperature.

\section{Discussion}

Our XAS measurements combined with DFT+$U$ calculations clearly illustrate the formation of two electronically distinct phases in hole-doped YTO, one with a hole-density $p = 0$ that resembles the $x = 0$ Mott insulating state, and the other with $p \sim 0.5$ that resembles the $x = 0.5$ metallic state. The finding clearly demonstrates the first-order nature of the IMT. As noted in the introduction, structural phase-separation into orthorhombic and monoclinic domains at low temperature is known to occur for Y$_{1-x}$Ca$_{x}$TiO$_{3}$ in the doping range $x = 0.37-0.41$ \cite{Kato2002a,Tsurui2004,German2018}. This is likely the result of large strains associated with the electronic phase separation. Note that the \textit{electronic} phase separation into hole-rich and hole-poor regions reported here should not be confused with \textit{chemical} phase separation into Ca-rich and Ca-poor regions; we are not aware of any evidence for the latter.

The question still remains, however, as to why the electronic phase separation in Y$_{1-x}$Ca$_{x}$TiO$_{3}$ occurs to a metallic-phase with such a large hole-concentration of $p^* \sim 0.5$ compared to the more modest prediction of $p^* = 0.27$ in ref. \cite{Yee2015}. 
Understanding this will likely require the consideration of the effects of electron-lattice coupling \cite{Kurdestany2017}, which was not considered in ref. \cite{Yee2015}. In particular, the role of elastic strain due to cooperative lattice distortions in the IMT physics in transition-metal oxides has been highlighted in recent work \cite{Guzman-Verri2019}. Although the Ti$^{3+}$ ion is nominally Jahn-Teller active, Jahn-Teller distortions are small, and hence considered irrelevant for ground-state properties \cite{Keimer2000,Varignon2017}. However, doping-induced short-range inhomogeneities are expected to be present, and could potentially be relevant \cite{Guzman-Verri2019}. It is a distinct possibility that this could contribute to the discrepancy between the theoretical prediction of $p^* = 0.27$ and the experimental result of $p^* \sim 0.5$.

Additionally, intermediate spiral, antiferromagnetic and ferromagnetic metallic phases are predicted in \cite{Yee2015,Kurdestany2017}. Signatures of such a magnetic metallic phase have been reported for hole-doped antiferromagnetic RTiO$_{3}$ with R = La, Pr, Nd and Sm \cite{Katsufuji1997}. 
For YTO, however, the ferromagnetic state extends to only $x = 0.2$, whereas the metallic phase emerges at $x_c \sim 0.35$. 
This discrepancy likely arises from the fact that refs. \cite{Yee2015,Kurdestany2017} start with a Hubbard model without considering the orbital order that is known to exist in YTO. This orbital order is known to lead to a ferromagnetic insulating ground state, rather than the antiferromagnetic insulating ground state that is expected from the Hubbard model \cite{Mochizuki2004}. Importantly, we note that no anomaly appears in the XAS data at $x \sim 0.2$ where the ferromagnetism disappears, which indicates that the electronic phase separation that we find here is rather insensitive to the specific magnetic ground state of the system. 

Another important point that warrants consideration is that the metallicity appears only at $x = 0.35$, which corresponds to a $\rho \sim 70 \%$ concentration of metallic $p = 0.5$ regions. However, the three-dimensional site percolation threshold is known to be $\rho \sim 30\%$ for a simple cubic lattice \cite{Kirkpatrick1976}. 
This indicates that the hole-rich and hole-poor regions likely form patterns, causing a significant increase in the percolation threshold. This would also explain the absence of the intermediate magnetically-ordered precolative-metal phase predicted in \cite{Yee2015,Kurdestany2017}. Indeed, stripe-like metallic and insulating phases have been observed in optical microscopy experiments for $x = 0.37$, close to the IMT, with a spatial extent on the order of several tens of microns \cite{German2018}. An alternative, yet seemingly less likely interpretation could be that the electronic phase separation actually involves $p = 0$ and some $p^* > 0.5$, in which case, the volume fraction of the metallic $p = p^*$ phase in $x = 0.35$ would be much less than $\rho \sim 70\%$. Note that $p^*$ cannot be significantly larger than 0.5 as the resistivity already starts to increase at $x = 0.7$ ($p \sim 0.7$) relative to $x = 0.5$ ($p \sim 0.5$), and eventually converges to a band-insulator at $x = 1$ ($p \sim 1$) \cite{Tokura1993,Taguchi1993}.


Finally, we contrast our findings for the three-dimensional (Y,Ca)TiO$_3$ perovskite with the quasi-two-dimensional high-$T_c$ cuprate superconductors, arguably the most-extensively studied doped Mott (charge-transfer) insulators. In the cuprates, the insulating behavior at half-filling breaks down already at nonzero ($p \sim$ 1\% or less) hole-doping \cite{Lee2005,Barisic2013}. However, Mott and Fermi-liquid characteristics coexist \textit{locally} until pure Fermi-liquid behavior is observed at high ($p \sim$ 30\%) doping \cite{Meissner2011,Haase2012,Storey2012}, a remarkable feature of the cuprates that is likely related to the prominent role played by nanoscale structural and electronic correlations \cite{Krumhansl1992,Egami1994,Pelc2019}. Unlike the proposed theoretical picture for (Y,Ca)TiO$_3$ \cite{Yee2015,Kurdestany2017}, the electronic inhomogeneity in the cuprates appears to emerge from inherent structural inhomogeneity \cite{Pelc2021}.

\section{Conclusions}

In conclusion, we find that the hole-doping effect in Y$_{1-x}$Ca$_x$TO$_3$ is two-fold: (i) electronic phase separation into hole-poor regions with a hole-density $p = 0$ that correspond to the Mott-insulator at $x = 0$, and hole-rich regions with a hole density $p^* \sim 0.5$ that correspond to the metal at $x = x^* \sim 0.5$; (ii) the emergence of in-gap states near the Fermi level, which increase in spectral intensity with increasing $x$ and decreasing temperature. 
This spectral intensity presumably arises from the metallic $p^* \sim 0.5$ phase that ultimately percolates and gives rise to bulk metallicity. Our results reveal the inherent first-order nature of the insulator-metal transition. To the best of our knowledge, the present work is the first spectroscopic evidence for electronic phase separation in a rare-earth titanate, and it sets the stage for future experiments and theoretical work in these model systems and in transition metal oxides in general.


\section{Acknowledgments}
We thank Damjan Pelc for valuable comments on the manuscript and Chris Leighton for the use of sample-polishing equipment. The work at University of Minnesota was funded by the Department of Energy through the University of Minnesota Center for Quantum Materials, under DE-SC0016371. This research used resources of the Advanced Photon Source, a U.S. Department of Energy (DOE) Office of Science User Facility operated for the DOE Office of Science by Argonne National Laboratory under Contract No. DE-AC02-06CH11357. Part of this work was carried out in the University of Minnesota Characterization Facility, which receives partial support from NSF through the MRSEC program. 

\bibliography{XAS_CaDoped.bib}

\widetext
\clearpage

\begin{center}
\textbf{\large Supplemental Figures}
\end{center}
\setcounter{equation}{0}
\setcounter{figure}{0}
\setcounter{table}{0}
\setcounter{page}{1}
\makeatletter
\renewcommand{\theequation}{S\arabic{equation}}
\renewcommand{\thefigure}{S\arabic{figure}}
\renewcommand{\bibnumfmt}[1]{[S#1]}
\renewcommand{\citenumfont}[1]{S#1}

\begin{figure}
\includegraphics[width=0.5\textwidth]{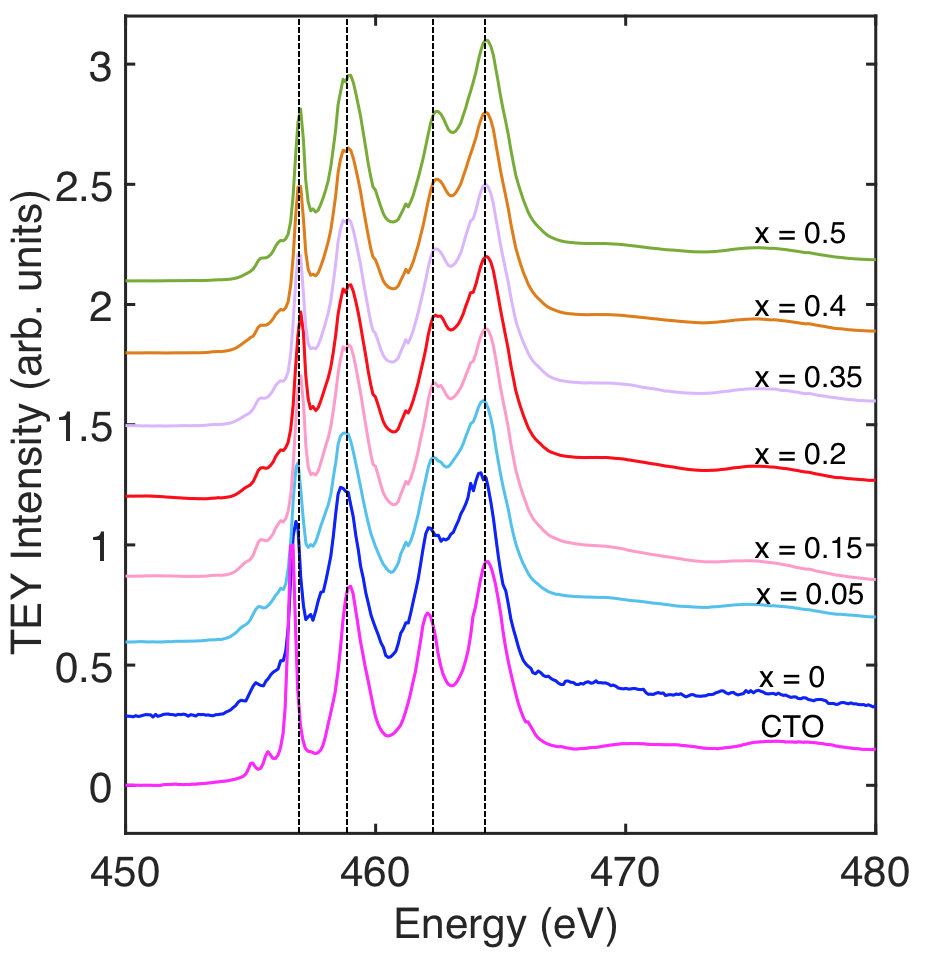}
\caption{(a) Ti $L$-edge XAS spectra obtained in the surface-sensitive TEY mode for different doping levels in Y$_{1-x}$Ca$_{x}$TiO$_{3}$, including reference spectra for CTO ($x=1$). The peaks in the spectra match those of CTO, which indicates that the Ti$^{3+}$ on the surface has mostly been oxidized to Ti$^{4+}$. The data were obtained at 15 K.}
\label{fig:TiL_TEY}
\end{figure}

\begin{figure*}
\includegraphics[width=0.9\textwidth]{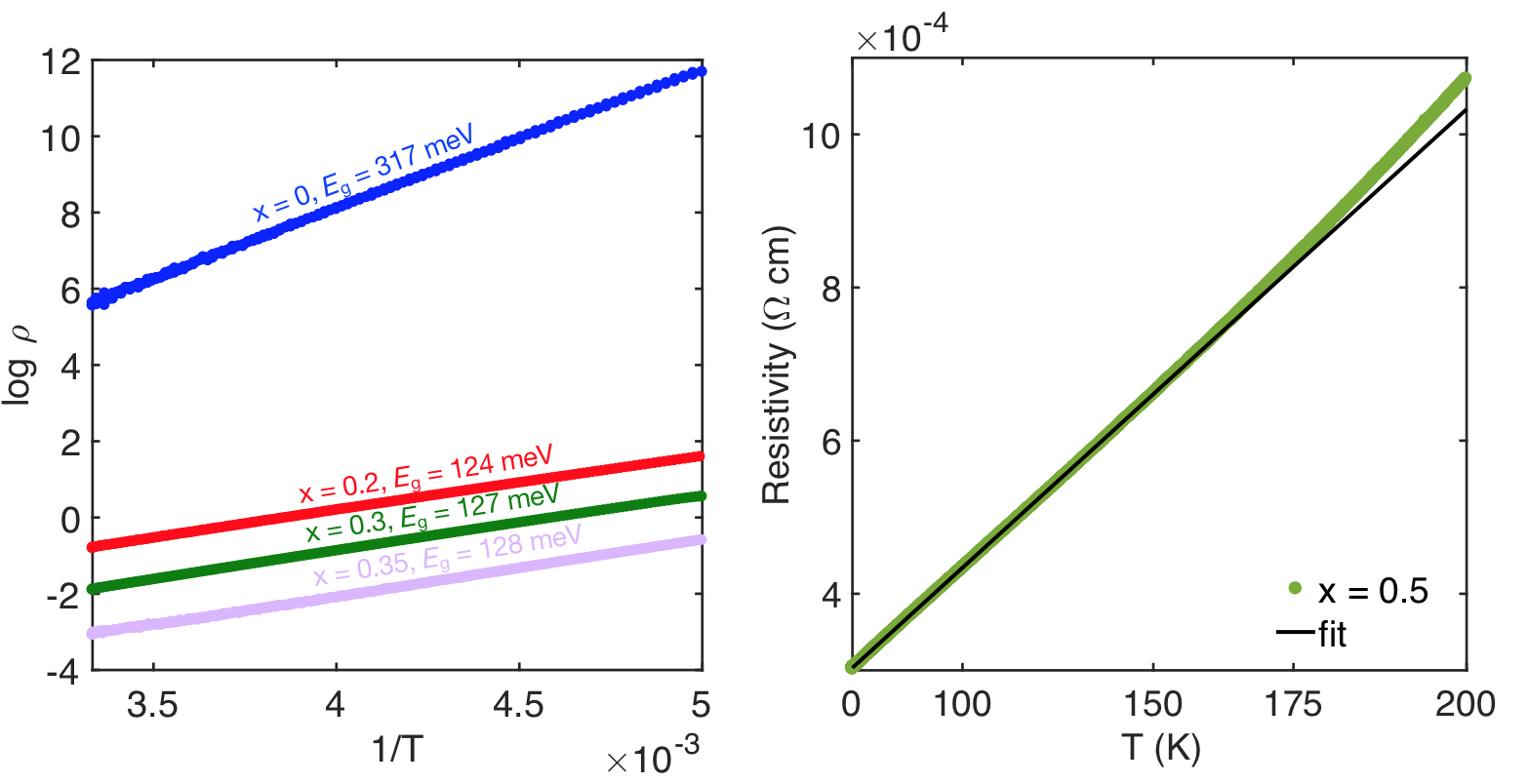}
\caption{(a) Activated resistivity $\sim e^{-E_g/k_B T}$ in the insulating regime in Y$_{1-x}$Ca$_{x}$TiO$_{3}$. (b) Power-law of the form $\sim T^{p}$ with $p = 2.48(3)$ observed in the metallic $x = 0.5$. The scale of the horizontal axis is $T^p$. The high temperature cut-off for the fit-range was varied between 75 and 160 K to estimate the error-bar for the exponent $p$.}
\label{fig:Res_Analysis}
\end{figure*}

\begin{figure}
\includegraphics[width=0.5\textwidth]{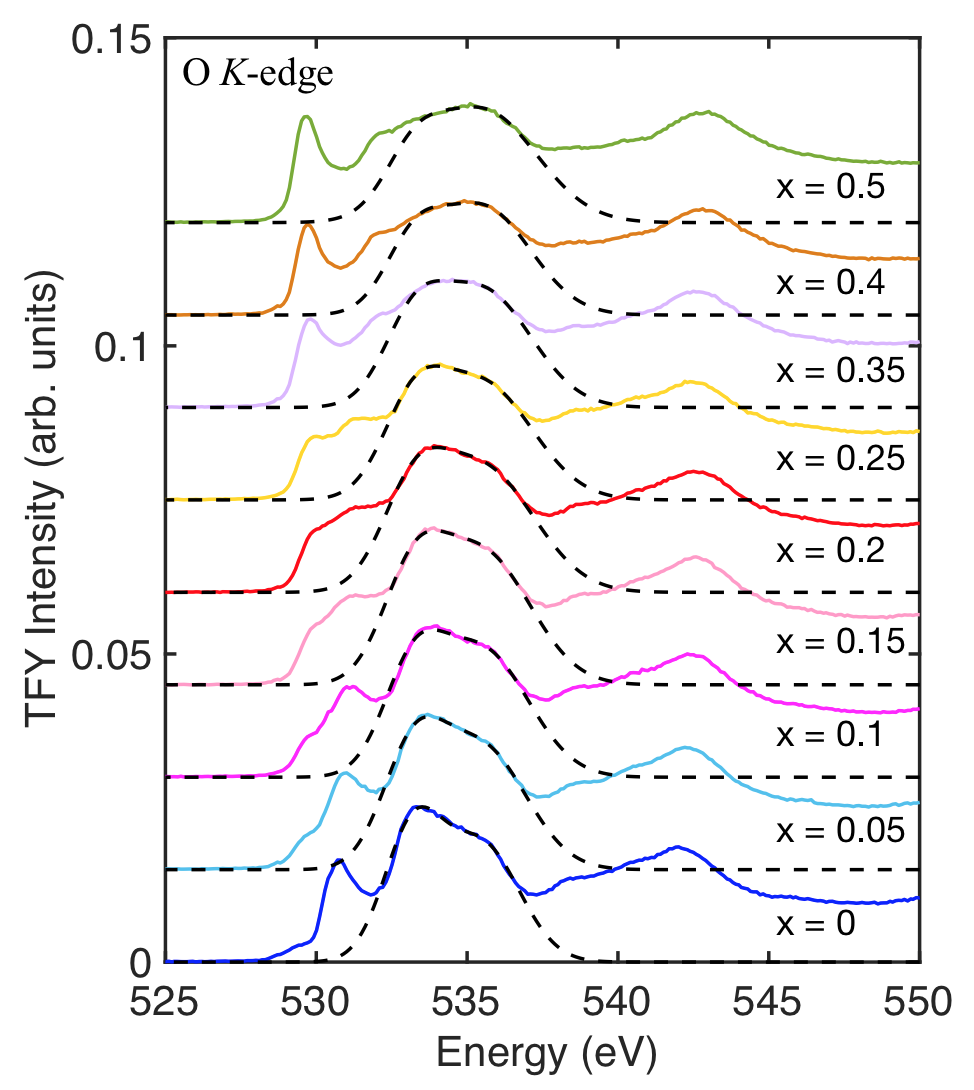}
\caption{Subtraction employed to extract the pre-edge region in the O $K$-edge XAS spectra at different doping levels in Fig.~\ref{fig:XrayOK}. Dashed black lines are the heuristic two-gaussian fits, with the peak positions fixed at 533.2 eV and 535.5 eV (based on the best fit for $x = 0$). Clearly, the fits capture well the data in this intermediate energy range. All data were taken at 15 K.}
\label{fig:OK_BG}
\end{figure}

\begin{figure}
\includegraphics[width=\textwidth]{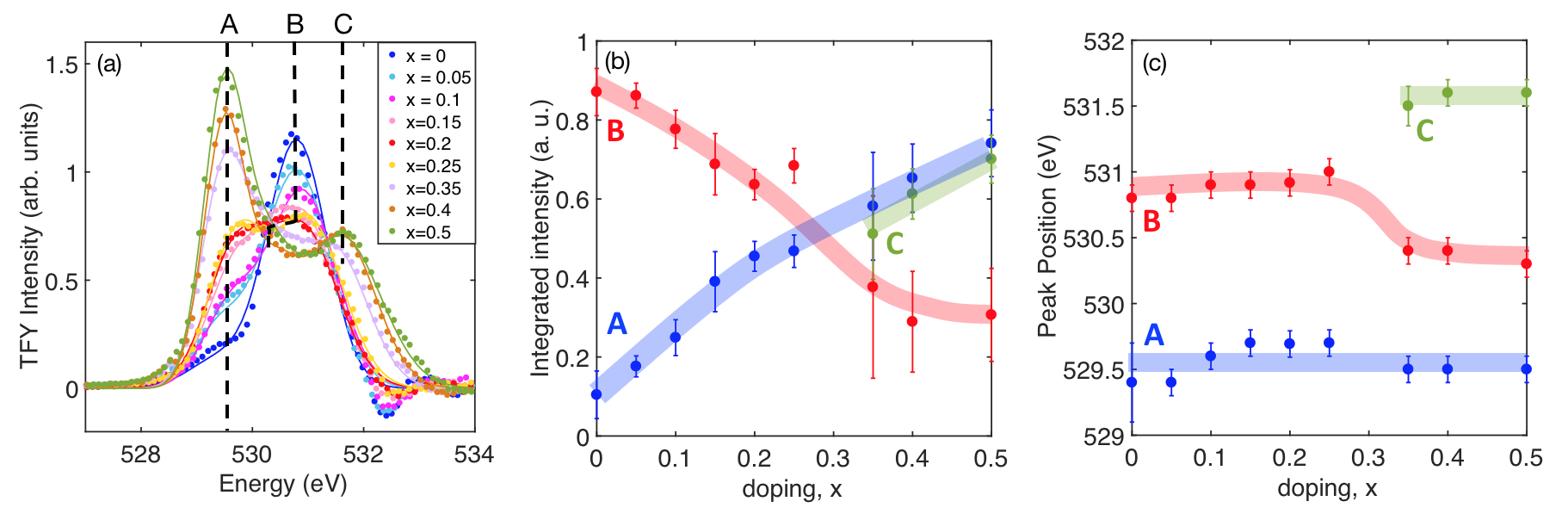}
\caption{(a) Pre-edge region in the O $K$-edge XAS spectra, obtained after normalizing the data to the intensity at 560 eV and employing the subtraction as in Figs.~\ref{fig:XrayOK} and~\ref{fig:OK_BG}. The lines are the results of fits to two or three gaussians, as described in the text. The vertical black dashed lines track the distinct peak positions. (b,c) Energy-integrated intensities and peak positions, respectively, of the gaussian peaks in (a). The qualitative results for the peak positions and integrated intensities obtained in Fig.~\ref{fig:XrayOK}(c,d) are seen to remain unaffected by the normalization.}
\label{fig:OK_BG_Norm}
\end{figure}

\begin{figure}
\includegraphics[width=0.8\textwidth]{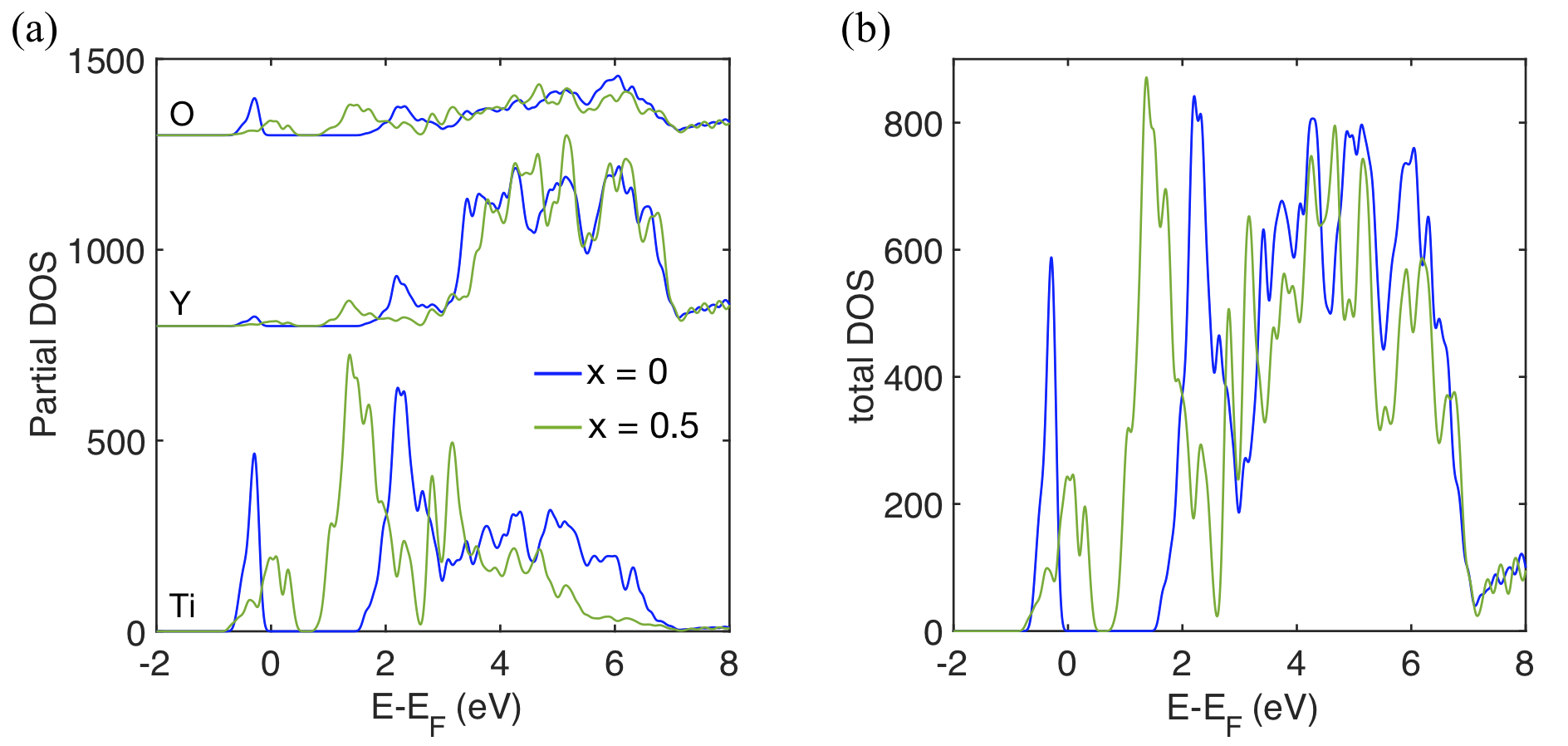}
\caption{Element-wise partial density of states (PDOS) and the total density of states (DOS) for $x = 0$ ($p = 0$) and $x = 0.5$ ($p = 0.5$), obtained from DFT+U calculations. The result of the calculation is convolved with a gaussian function of width 0.1 eV (FWHM) to mimic the experimental resolution. The qualitative changes in the total DOS are clearly seen to be reflected in the Ti PDOS. We therefore compare the O $K$-edge data with the Ti PDOS. The high-energy region ($>3.5$ eV) shows no observable shifts, which justifies the use of fixed peak positions of the two gaussians used to capture the spectral intensities above the pre-edge region in Fig.~\ref{fig:XrayOK}.}
\label{fig:DOSth}
\end{figure}

\begin{figure}
\includegraphics[width=\textwidth]{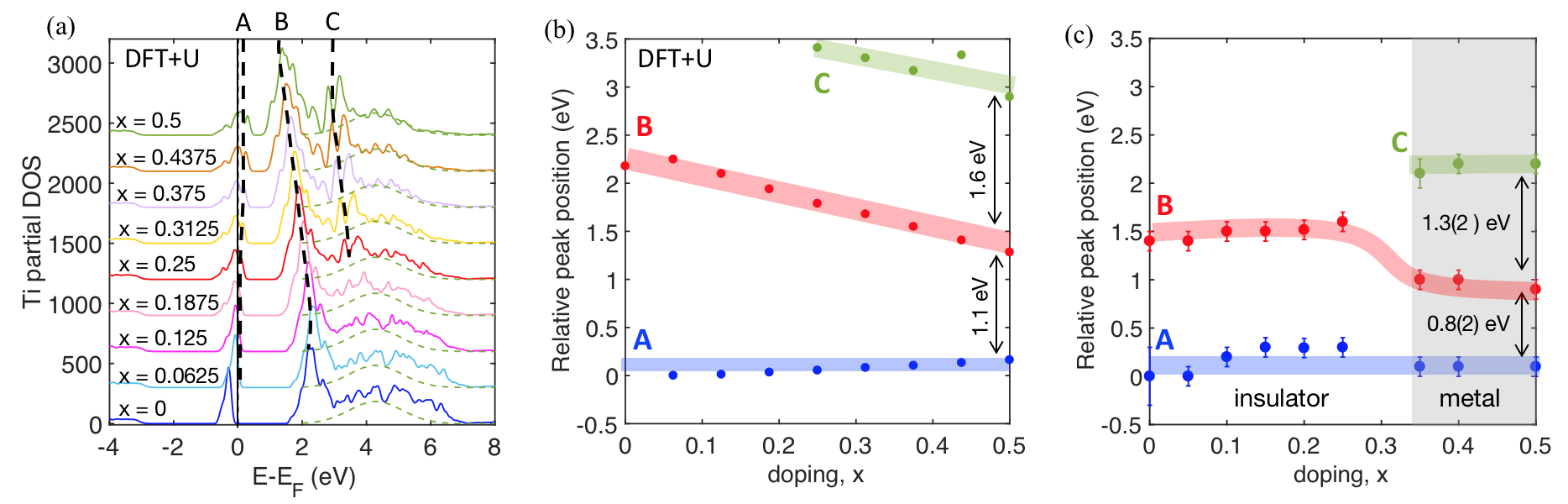}
\caption{(a) Doping dependence of the Ti PDOS obtained from DFT+U calculations, without considering any phase electronic separation. The result of the calculation is convolved with a gaussian function of width 0.1 eV (FWHM) to mimic the experimental resolution. The black dashed lines track the three distinct peaks that correspond to the experimental pre-edge peaks in Fig.~\ref{fig:XrayOK}(b). The dashed green lines are a gaussian fit to the broad high-energy component at $x = 0.5$, used to highlight the transfer of PDOS from the higher-energy region to the lower-energy region, forming peak C. The doping dependence of the peak positions are displayed in (b). Note that the peak position for A is chosen as the midpoint of the nonzero-PDOS region just above the Fermi level, as XAS only discerns states above the Fermi level. The doping dependence of the experimental peak positions from Fig.~\ref{fig:XrayOK}(f) is repeated in (c) for ease of comparison. Whereas the experimentally-obtained peak B position shows a discontinuity between $x = 0.25$ and 0.35, the theoretical peak position exhibits a qualitatively different, continuous change. Note that we have assumed $p = x$ here at all dopings, in displaying the DOS.}
\label{fig:DOSNoPS}
\end{figure}

\begin{figure}
\includegraphics[width=0.7\textwidth]{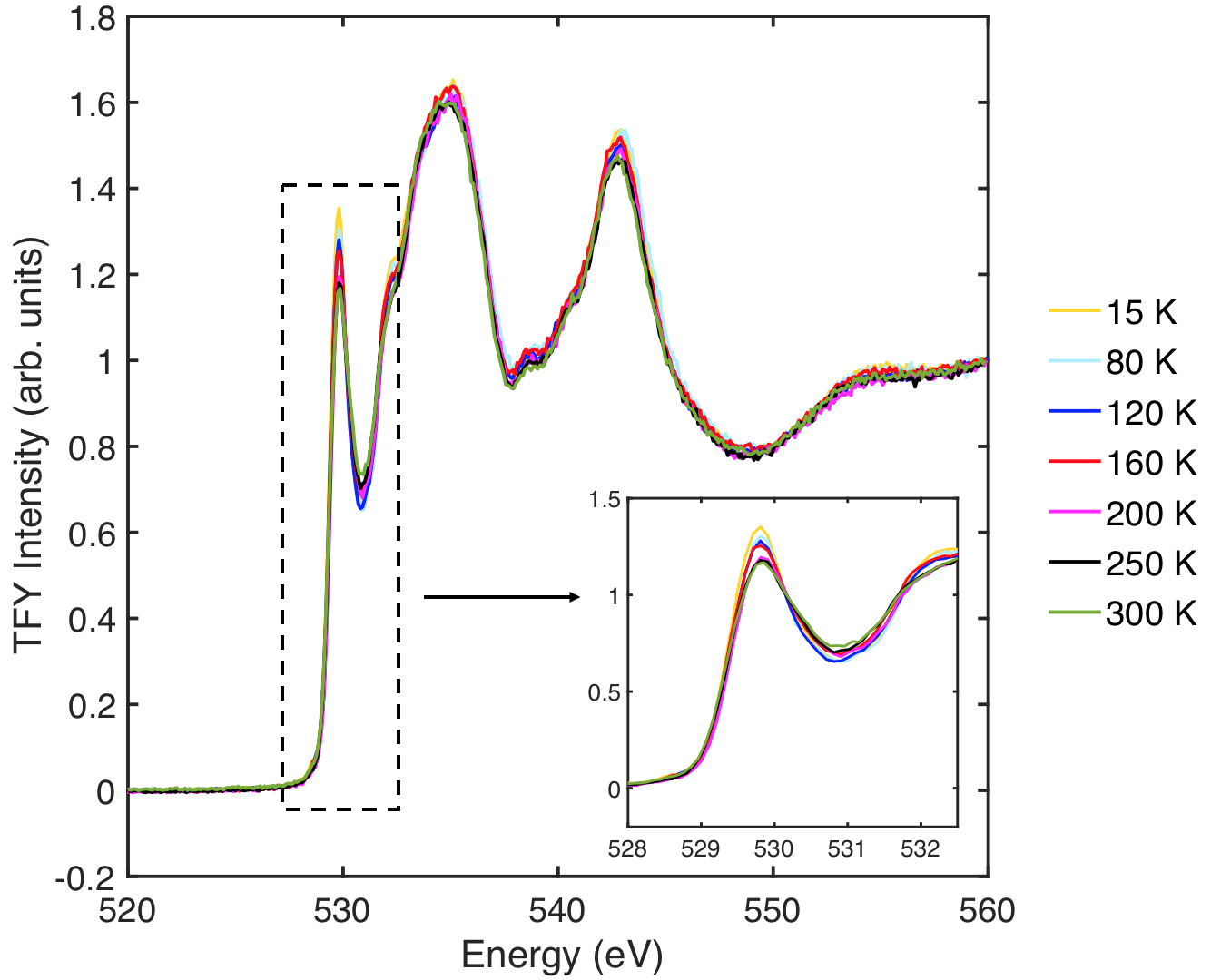}
\caption{O $K$-edge XAS spectra measured in the TFY mode for $x = 0.4$ at a number of temperatures. The data were obtained on heating. The inset shows a magnified view of the pre-edge region. The spectra are normalized to the intensity at 560 eV.}
\label{fig:OK_TFY40}
\end{figure}

\begin{figure}
\includegraphics[width=0.45\textwidth]{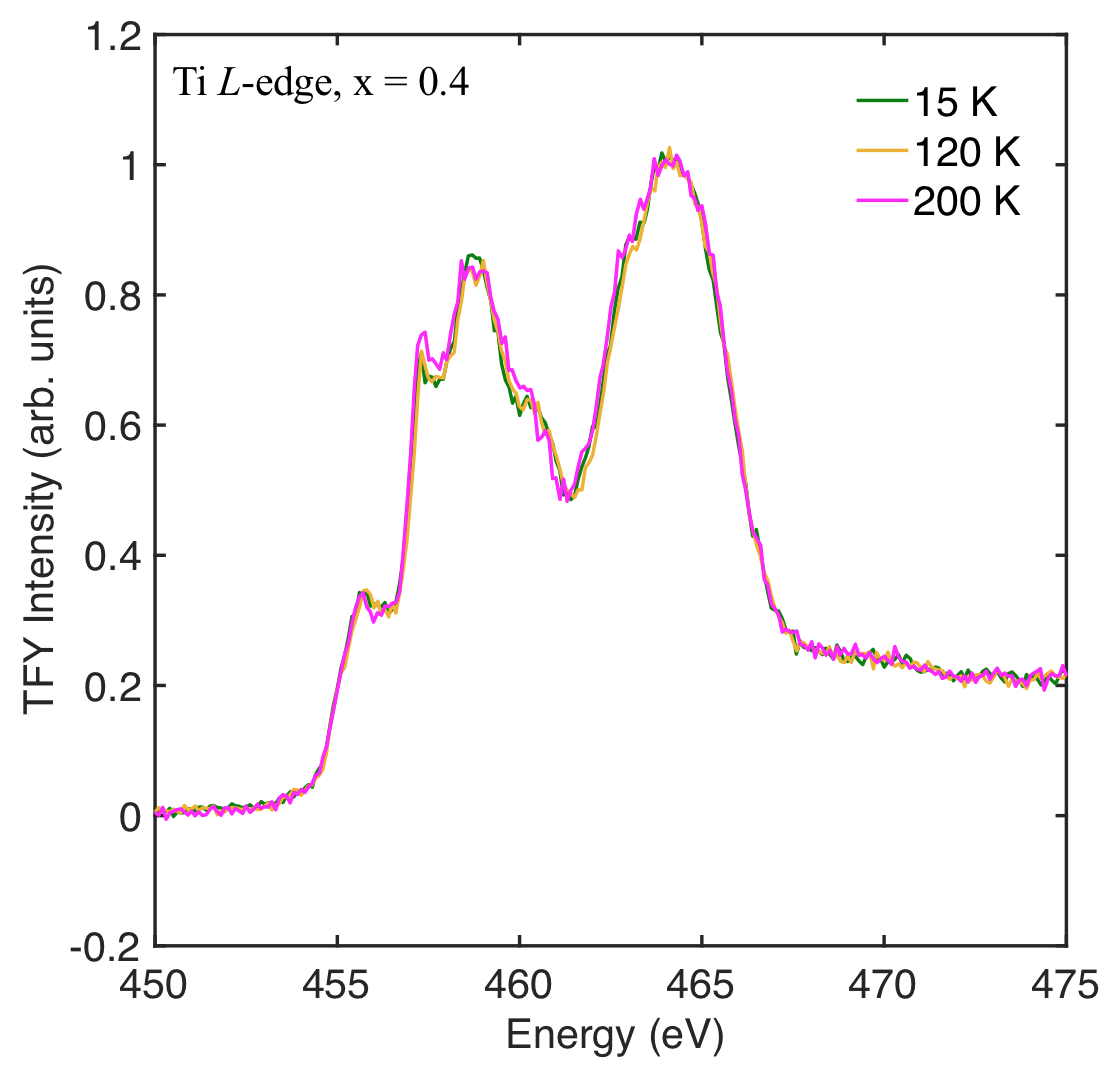}
\caption{Ti $L$-edge XAS spectra measured in the TFY mode for $x = 0.4$ at a different temperatures. The data were obtained on heating. The background intensity at 450 eV has been subtracted and the spectra are normalized to the highest peak intensity.}
\label{fig:Ti_TFY40}
\end{figure}

\begin{figure}
\includegraphics[width=0.9\textwidth]{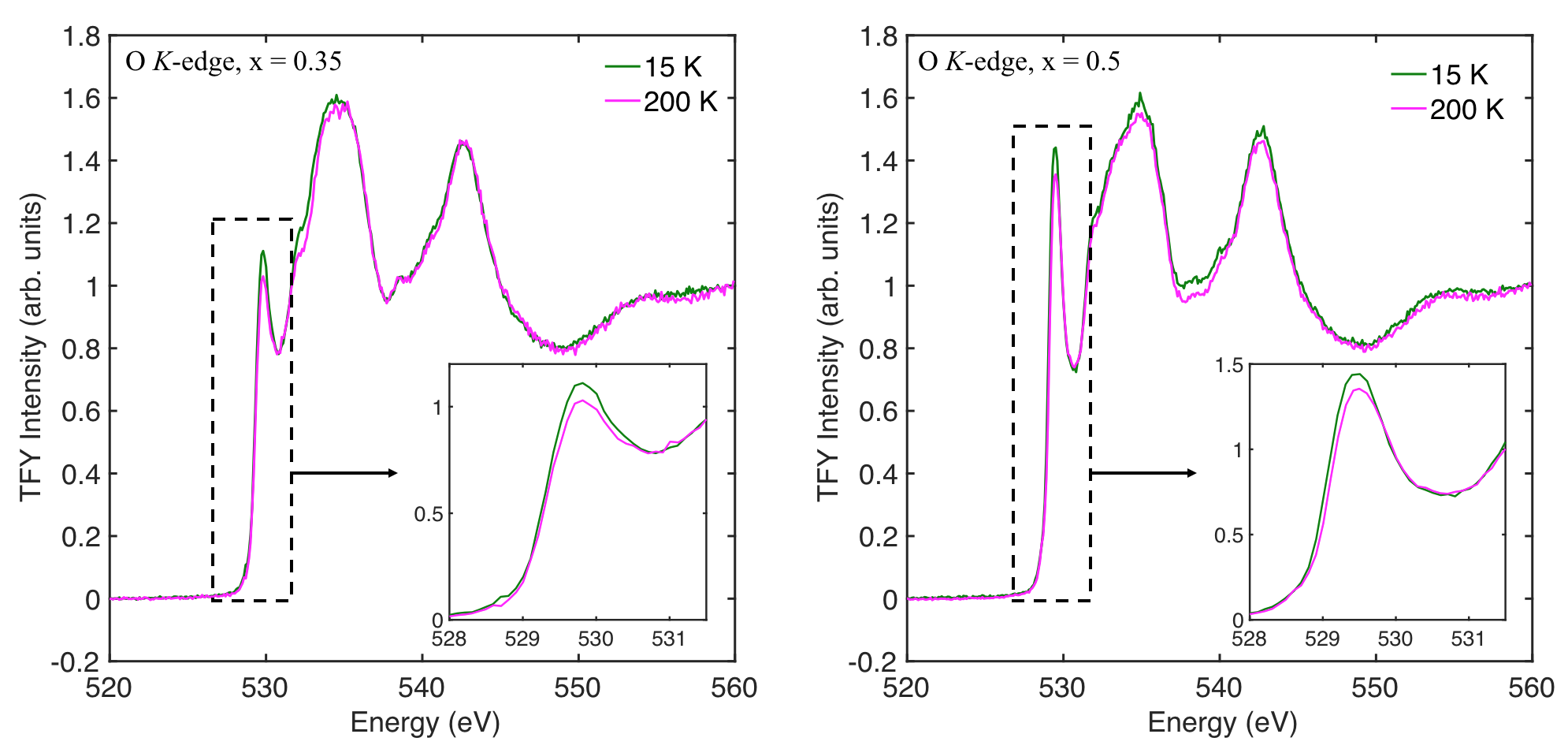}
\caption{O $K$-edge XAS spectra measured in the TFY mode for $x = 0.35$ and $x = 0.5$ at a number of temperatures. The inset shows a magnified view of the pre-edge region. The spectra are normalized to the intensity at 560 eV.}
\label{fig:OK_TFY}
\end{figure}

Here we document the Ti $L$-edge XAS spectra in TEY mode (Fig.~\ref{fig:TiL_TEY}), the activated behavior of the resistivity in insulating samples $x = 0-0.3$ (Fig.~\ref{fig:Res_Analysis} (a)), the low-temperature power-law temperature-dependence of the resistivity in the metallic $x = 0.5$ sample (Fig.~\ref{fig:Res_Analysis} (b)), the subtraction used to extract the pre-edge region of the O $K$-edge XAS (Fig.~\ref{fig:OK_BG}), the insensitivity of the qualitative results in Fig.~\ref{fig:XrayOK}(c,d) to normalization of spectra (Fig.~\ref{fig:OK_BG_Norm}), the doping dependence of the element-wise partial density of states and the total density of states obtained from DFT+U calculations (Fig.~\ref{fig:DOSth}), the qualitative difference between the experimental results and the doping dependence of the Ti partial density of states obtained directly from DFT+$U$ calculations without considering any electronic phase separation (Fig.~\ref{fig:DOSNoPS}), the temperature dependence of the O $K$-edge XAS spectra in TFY mode at $x = 0.4$ for additional temperatures apart from those in Fig.~\ref{fig:XrayOKTdep} (Fig.~\ref{fig:OK_TFY40}), the temperature dependence of the Ti $L$-edge XAS spectra in TFY mode for $x = 0.4$ (Fig.~\ref{fig:Ti_TFY40}), and the temperature dependence of the O $K$-edge XAS spectra in TFY mode for $x = 0.35$ and $x = 0.5$ (Fig.~\ref{fig:OK_TFY}).

\clearpage

\end{document}